\documentclass[letterpaper, 10 pt, conference]{ieeeconf}  

\IEEEoverridecommandlockouts                              
\usepackage[utf8]{inputenc}
\usepackage[T1]{fontenc}
\usepackage{amsmath}
\usepackage{comment}
\usepackage{verbatim}  
\usepackage{graphicx}
\usepackage{amsmath,amssymb,amsfonts}
\usepackage{mathtools}
\usepackage[super]{nth}
\usepackage{pdfpages}
\usepackage{algorithmicx}
\usepackage[ruled,vlined,linesnumbered]{algorithm2e}

\title{\LARGE \bf
Fundamentals of Next-generation Network Planning
}

\author{M. Umar Khan$^{1}$ 
\thanks{*This work was not supported by any organization}
\thanks{$^{1}$M. Umar Khan is with Faculty of Computer Engineering, COMSATS University Islamabad, Pakistan
        {\tt\small umar\_khan@comsats.edu.pk}}%
\thanks{}%
}

\begin{document}

\maketitle
\thispagestyle{empty}
\pagestyle{empty}

\begin{abstract}

The fifth-generation (5G) of cellular communications is expected to be deployed in the next years to support a wide range of services with different demands of peak data rates, latency and quality of experience (QoE). To support higher data rates and latency requirements third-generation partnership project (3GPP) has introduced numerology and bandwidth parts (BWPs), via new radio (NR) for service-tailored resource allocation. Legacy 4G networks have generated extensive data, which combined with crowd-sourced LTE infrastructure insights, enables identification of high-traffic 5G deployment area (5GDA) for planning new services. Given the mission-critical nature of 5G services, QoE is a big challenge for MNOs to guarantee peak
data rates for a defined percentage of time. This work studies the fundamentals of 5G network planning methods that reconciles coverage-capacity trade-offs through balanced radio network dimensioning (RND), leveraging pragmatic NR modeling, and data-driven strategies to minimize deployment costs and reduce cost-per-bit.  
\end{abstract}

\section{INTRODUCTION}

The first generation (1G) of cellular network was based on analog protocols in order to provide basic voice service to the subscribers. Next, the second generations (2G) known as global system for mobile (GSM) was introduced as a successor of 1G, and fundamentally addressed as a circuit-switched digital network. It was designed on first digital standards of time division multiple access (TDMA) or GSM for improved coverage. The third generation (3G) replaced the 2G with universal mobile telecommunication system (UMTS). The 3G was designed to provide voice, internet and first mobile broadband with data considerations through high speed packet access (HSPA) to enhance the capacity. To further enhance the capacity, the fourth generation (4G) was introduced which was designed primarily for data based on long term evolution (LTE) protocols to provide true mobile broadband. To that end, legacy cellular generations were fundamentally designed and planned to improve coverage and capacity. However, the fifth generation (5G) is different which demands several use cases as a service to provide ultra low latency communication (URLLC), massive machine type communication (mMTC) and enhanced mobile broadband (eMBB). To fulfill the requirements of 5G services, the 3GPP has proposed new radio (NR) based on numerology and bandwidth parts (BWPs). Consequently, to offer new services based on 5G NR, the new service requirements must be an integral part of the network design and planning.

\begin{figure}[h]
	\centering  
	\begin{center}           
		\includegraphics[width=.50\textwidth]{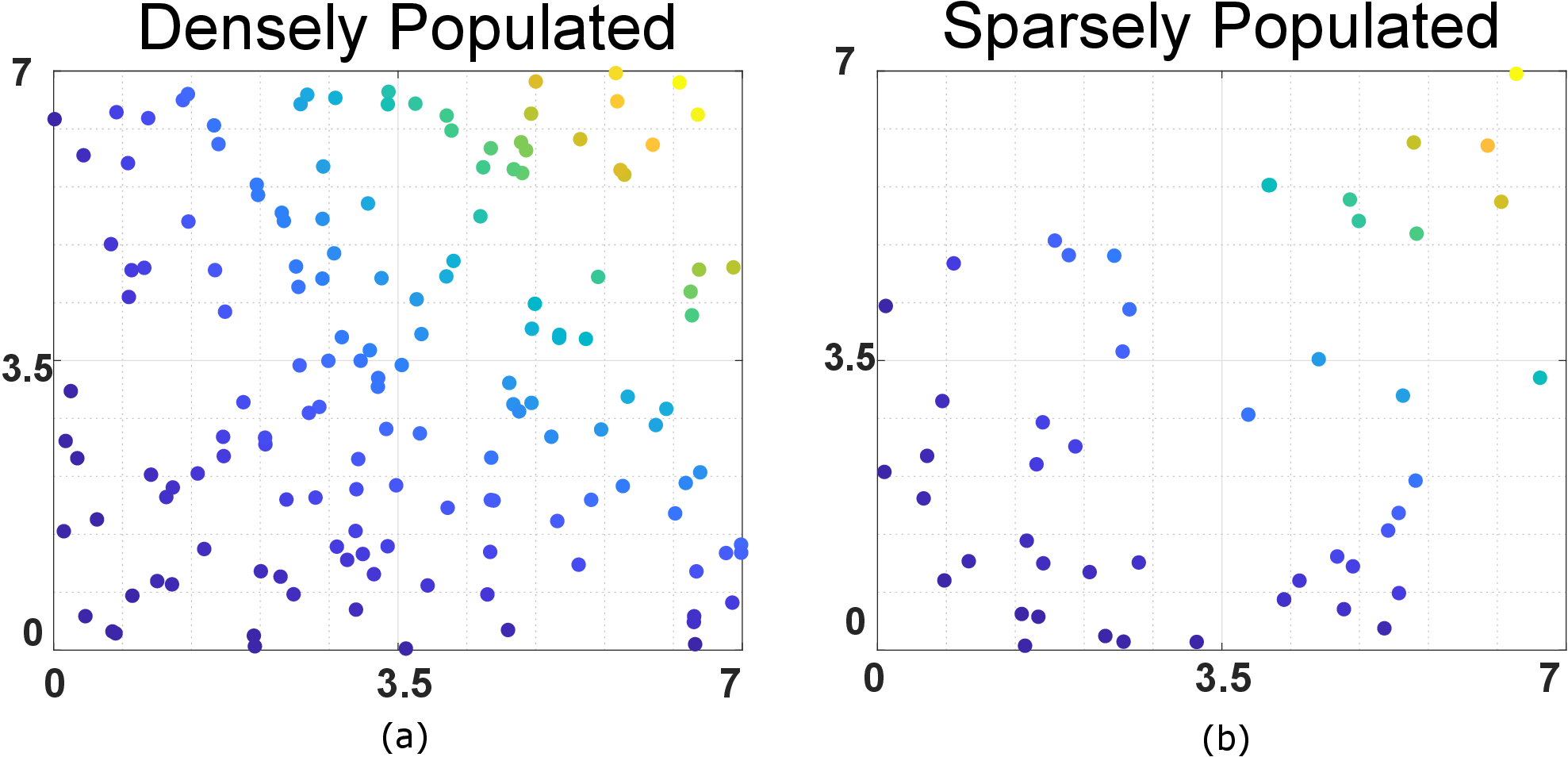}
	\end{center} 
	\caption{Example of subscriber density in $7\times 7 ~\text{km}^2$ identical areas. The densely populated and higher traffic area in figure (a) represents the region where subscriber's traffic density is higher compared to the sparsely populated subscribers area in figure (b) where traffic density is at lowest.}\label{fig:CBP_example}
\end{figure}

The objective to utilize network data is to extract traffic information and to identify the area of the highest traffic density to perform network dimensioning for deployment of new base stations. The network data helps MNO to reduce the cost per bit by achieving higher network utilization. Whereas the higher network utilization can only be accomplished if base stations are deployed in the highest traffic density region by extracting traffic information from the network data.  For instance, consider the two areas of $7 \times 7 \text{ km}^2$ given in Fig.\ref{fig:CBP_example} representing (a) densely populated area with higher traffic and (b) sparsely populated region with lower traffic density. 

The same number of base stations would cover both identical areas. However, the subscriber density and corresponding network traffic is different, this situation possibly leads to under utilization of the resources in the sparsely populated area (Fig.\ref{fig:CBP_example} (b)) where traffic density is lower. Assuming that the capital expenditure (CAPEX) and operational expenditure (OPEX) of base station deployment are constant for both areas, the lower network utilization brings a higher cost per bit for the MNO. On the other hand, in the case of a densely populated area (Fig.\ref{fig:CBP_example} (a)) the network utilization will be higher, leading to a lower cost per bit as compared with the former case. The densely populated region of the highest traffic density can be identified using the LTE location identifiers assisted by network data. In mobile network, the geography is divided into several tracking areas marked by different codes or IDs corresponding to LTE identifiers. Therefore, to identify certain area and traffic information, combinations of LTE identifiers and corresponding IDs can be used for network data acquisition.

\section{Related Work}

Network planning has originated a significant body of research (see \cite{MonaJou,Ataufiq} and references therein) for LTE networks, though it faces a number of challenges regarding the deployment of 5G networks, including the self-organizing capability of 5G networks \cite{perez2016}, broadcast and multicast in smart grids \cite{saxena2017} and ultra-dense topologies \cite{tseng2015,udn1}. Traditionally, cell planning has followed a human-centric approach \cite{humancentric,humancentric1,humancentric2}, leading to network deployments where the number and location of base stations satisfy the traffic demanded by applications such as video streaming, Voice over IP (VoIP) and social networks. However, the needs of new machine-centric applications and scenarios (e.g., smart cities and emergency alerts) bring to 5G networks very stringent requirements in the form of strict delays or ultra-reliable links. Therefore, the proposals for 5G RNP should explore how to deal with different types of services simultaneously. The use of numerology and BWPs endows 5G NR \cite{TS38300} with a degree of flexibility so that the radio interface is configured to accommodate such diversity of services. An important aspect of telecommunication networks is the quality of service (QoS) \cite{qoslte}, which has been taken into consideration in mobile networks under very different forms, from guaranteeing a minimum data rate \cite{khasef2016} to throughput and delay constraints \cite{maule2018}. However, 5G services and applications require more than ensuring a given value of a conventional communications metric, leading to a broad adoption of the concept of QoE introduced in LTE networks \cite{qoe1,qoe2,qoe3}. Therefore, we study the probabilistic modeling of the 5G radio resource control (RRC) states to ensure the availability of the peak data rates for a required percentage of time.  

The majority of studies on radio network planning make use of forecast data, usually estimated traffic demand or traffic modeling, to design the network deployment (see for instance \cite{qos2,Yang2016,Gonzalez2016,Muhnoz2018}). As an alternative, there exist organizations that collect real mobile network data and make these data available on line, both in raw and processed form, making a more informed interpretation and decision-making possible. In this study, we focus on utilizing the available network data for 5G NR network planning based on algorithms and learning techniques, as these data can help us achieve a more efficient network planning design, especially in terms of topology and cost. The current 4G cellular networks are generating a huge amount of network data which are stored and processed by different entities for commercial usage \cite{opensignal,nperf}. These data can be utilized with ML-based approaches to enhance the network performance in terms of techno-economic aspects. However, these databases are expensive and the network data are sensitive information of the MNOs that cannot be made available publicly. Consequently, we resource to crowdsourced open data to leverage the existing infrastructure from OpenCelliD by Unwired Labs \cite{opencellid}. It is a project where community collaborates in collecting GPS coordinates of cell towers and their corresponding location identifiers and make it available for academic research. The data acquired from the OpenCelliD can be utilized to identify the highest traffic density area per \text{${km}^2$}. Besides, these data can aid the learning techniques to reveal insights of the densest traffic area and corresponding legacy infrastructure on the cluster level. Therefore, meaningful traffic information obtained from these data can assist in the planning of the access network where new 5G gNBs will be deployed.   

On the other hand, 5G does not merely bring challenge to the access network but backhaul network also requires consideration as new gNBs will be connected to the core network through the backhaul nodes. In this context, the data rate and latency requirement of the gNBs impact the backhaul directly, such that the backhaul should bare the high capacity demands with required latency and minimum cost \cite{monaBH,monaBH1}. The knowledge of the existing macro-sites and the available capacity of the nodes can be used to design the backhaul network topology to provide minimum cost connectivity to the gNBs. In this context, a mathematical model can be formulated for the backhaul connectivity planning to minimize the deployment cost such that the capacity constraint is fulfilled.

\section{Radio Network Planning}
The process of provisioning, positioning and computing the necessary number of base stations along with the network parameters to deliver adequate coverage, capacity and quality of service (QoS) through the cellular network is known as radio network planning (RNP). The three fundamental objectives; coverage, capacity and QoS are usually approached simultaneously in RNP exercise. Radio network dimensioning (RND) is the primary phase of the cellular network planning that aims to accommodate the necessary coverage and capacity in the network to be deployed in the target area. RND is fundamentally focused on the link budget analysis (LBA) refers to coverage dimensioning. The LBA considers a suitable propagation model to compute the maximum coverage range, which is used to estimate the number of required radio sites in the deployment area of interest. Moreover, in the RND phase the subscribers are assumed to be uniformly distributed in the area where all the BSs are evenly loaded. At the same time, capacity dimensioning is performed to determine the cell range based on the number of subscribers to be served by the base station. This reasonable approach enables the planning engineers to perform a quick analysis on coverage and capacity cell range which consequently determines the required number of radio sites in the deployment area. Once the primary phase is completed, the second phase of the RNP deals with the frequency planning, mobility parameters and antenna azimuth. Finally, the third phase deals with the optimization and fine tuning of the corresponding parameters based on drive tests, service utilization and real user distribution in the deployment area \cite{MonaJou}. In this thesis, we study the primary phase 'RND' with focus on service-based dimensioning assisted by real network data with the objective to achieve a balanced network design.  

\begin{figure}[h]
	\centering
	\includegraphics[width=.50\textwidth]{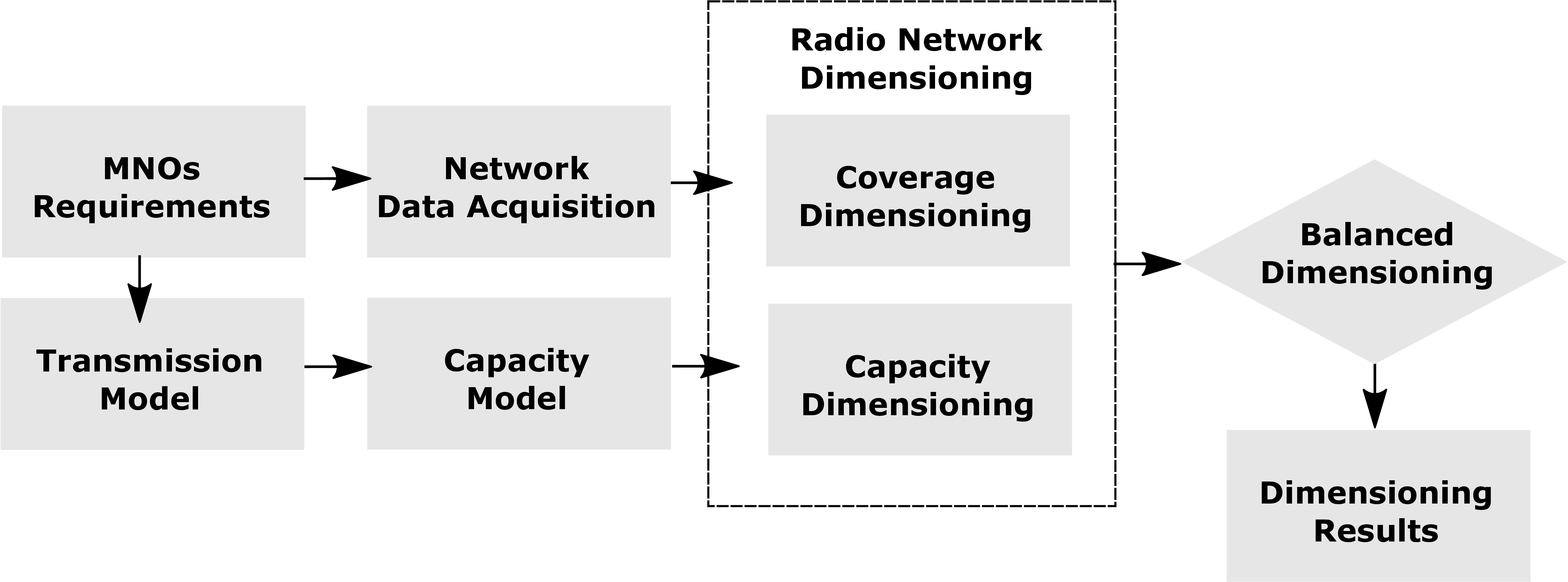}
	\caption{Necessary blocks for RND process to provide balanced dimensioning} \label{Fig:RND_basics}
\end{figure}

The RND phase refers to perform coverage and capacity dimensioning to estimate different parameters such as appropriate transmission power, cell range and required number of radio sites in the target area. This phase is dependent on different necessary inputs such as mobile network operators (MNOs) requirements, network data acquisition, transmission and capacity models as shown in Fig.\ref{Fig:RND_basics}. The MNOs' requirements are reflected in transmission model and in network data acquisition entities. In transmission model, physical layer parameters of offered services are defined which also translates into the capacity of the base station refers to the capacity model entity of Fig.\ref{Fig:RND_basics}. On the other hand, MNOs' requirements of selecting the highest traffic region and limiting the size of planning area as a financial constraint are considered through network data acquisition entity. The demarcated area for deployment is provided as an input to coverage dimensioning to perform LBA according to the physical layer parameters and assumed cell load to determine cell range and number of required radio sites for coverage. Whereas, capacity dimensioning is based solely on the transmission and capacity model to determine cell range and number of required radio sites for capacity. In capacity dimensioning, actual cell load is determined based on the subscriber density in the target area.        

\begin{figure}[h]
	\centering
	\includegraphics[width=.450\textwidth]{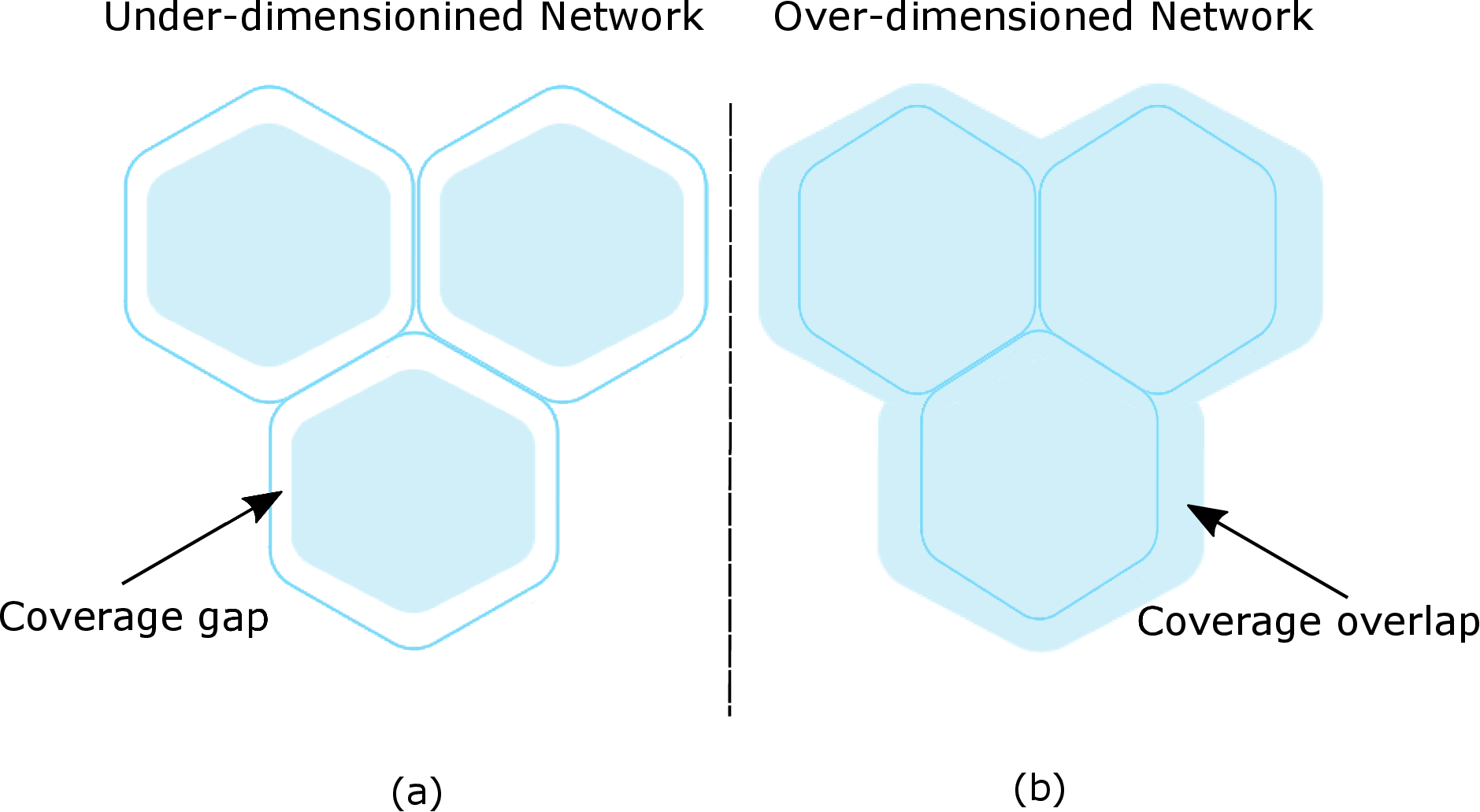}
	\caption{The solid boundaries represent the cell range obtained from the coverage dimensioning whereas the blue hexagons represent the cell range from the capacity dimensioning. The cells in figure (a) shows an under-dimensioned network whereas (b) represents an over-dimensioned network.} \label{balanced_network}
\end{figure}

The balanced dimensioning entity in Fig.\ref{Fig:RND_basics} refers to compare the coverage and capacity dimensioning results of cell range and cell load. The coverage targets refer to offer a certain throughput to subscriber at the cell edge which determines the corresponding cell range, the coverage cell range is represented by solid boundaries in Fig.\ref{balanced_network}.  The capacity targets refer to the ability of a base station to serve a certain number of subscribers which corresponds to a specific capacity cell range, the capacity cell range is represented by blue hexagons in Fig.\ref{balanced_network}. The coverage and capacity cell range are the mathematical tools for the estimation of the appropriate cell range and to determine if the network design is balanced or not \cite{MonaJou}. If the coverage cell range value is larger than the capacity cell range then the network design will be under-dimensioned and coverage gaps will be observed as shown in Fig.\ref{balanced_network}(a). This means that the coverage is provided to a larger area whereas the capacity is insufficient for the corresponding area. This situation will lead to over-utilization of the deployed base stations as there is a mis match in the actual and assumed load of the base stations. On the other hand, if the capacity cell range is larger than the coverage cell range, the network designed will be over-dimensioned and coverage overlaps will be observed as shown in Fig.\ref{balanced_network}(b). In this case, the network will be under-utilized as the capacity cell range is offered in a larger area, though, the actual cell load or the number of subscribers in the target area is smaller. Therefore, it is important to provide a balanced network design in the RND phase which can be guaranteed if the difference in cell range and the cell loads are minimized under given thresholds defined by the MNO. Moreover, traffic information acquired from network data can assist the RND phase to perform planning in the highest traffic density area in order to achieve higher network utilization with minimum cost per bit. 

\subsection{Identification of Tracking Area}
\begin{figure}[h]
	\centering
	\includegraphics[width=.450\textwidth]{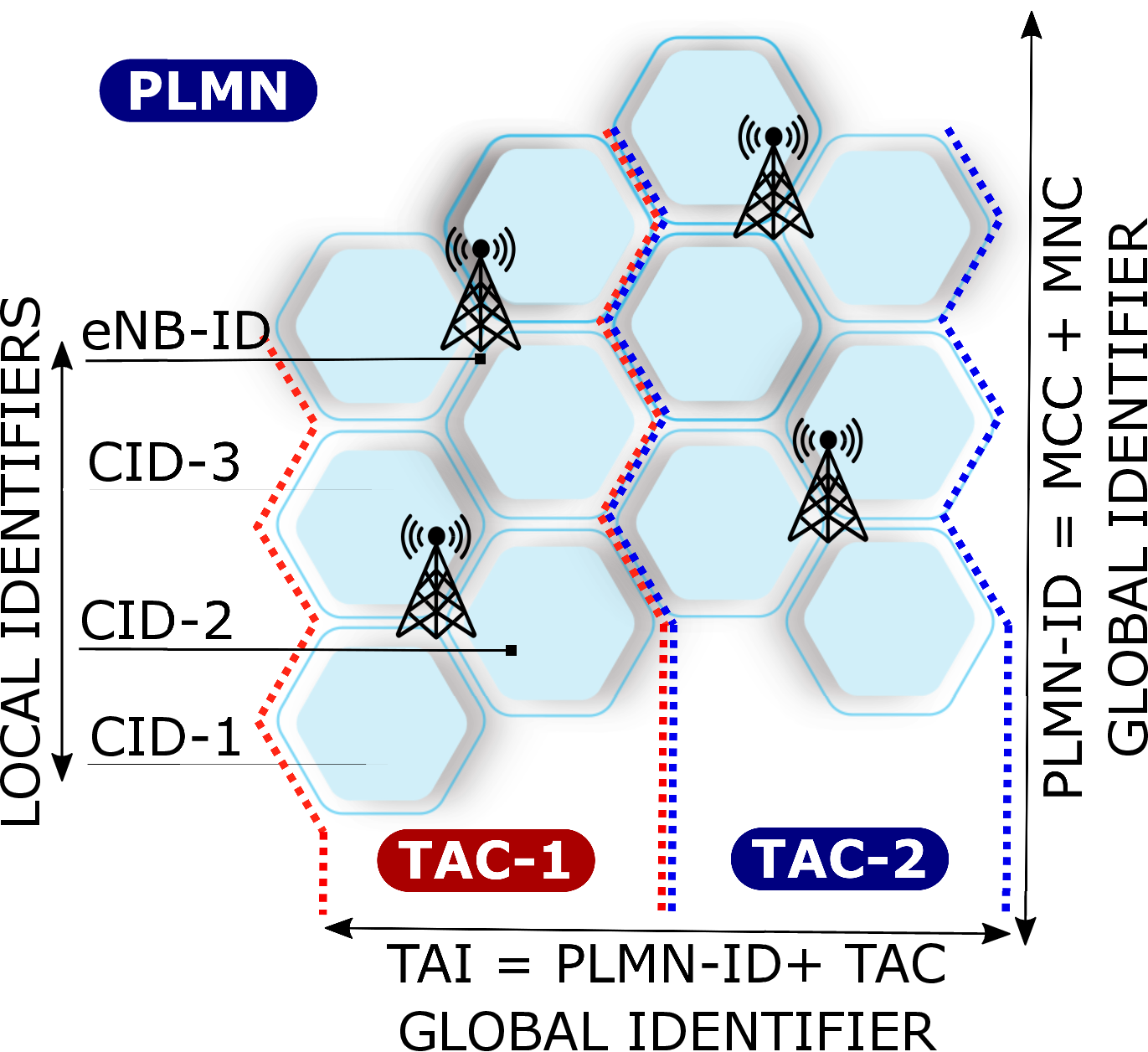}
	\caption{The standard LTE identifiers for tracking area identity (TAI) used to identify and process different variables in the network data} \label{lte_identifier}
\end{figure}

The tracking area (TA) in cellular networks is a logical grouping of existing cells and corresponding infrastructure to manage and identify the locations of both the user and the BS. The code that uniquely identifies the TA is the tracking area code (TAC), which represents a particular geographical area where BSs are located and can be uniquely identified. The TAC variable in the database has a significant importance in our study to identify the region of highest traffic. The TAC, along with other recorded variables in the network database, are the LTE identifiers as defined by 3GPP, which serve to provide unique and global identification throughout the network. These identifiers are classified into global and local identifiers as shown in Fig.\ref{lte_identifier}, respectively.

\subsection{Global Identifiers}
The global identifiers correspond to a globally unique identification of an area belonging to a particular MNO. In an area where multiple MNOs have installed their network infrastructure, a global identifier can be used to identify the network and the corresponding geographical area. For instance, consider the logical topology of the LTE cellular network inside the public land mobile network (PLMN) in  Fig.\ref{lte_identifier} with two TAs i.e., TAC-1 and TAC-2, which have unique codes inside the PLMN. The PLMN-ID represents a globally unique identifier of the network with the combination of mobile country code (MCC) and mobile network code (MNC). The tracking area ID (TAI) is defined as a globally unique identifier in the network for an area of a particular MNO corresponding to the combination PLMN-ID and TAC. 

\subsection{Local Identifiers}
The local identifiers refer to a unique identification of an area within the network infrastructure of an MNO.  Inside the network, identification of the installed base stations and the cells are provided by the identifiers. Each MNO has an unique identifier for each cell ID (CID) and each eNB (eNB-ID), as shown in Fig.\ref{lte_identifier}. The combination of TAC and CID can be used to identify a geographical region of interest and learn about the corresponding traffic in the MNO's network. 

\subsection{5G Numerology and BWPs}
5G new radio or NR refers to a new radio access technology designed and developed by 3GPP for the air interface of next generation of cellular networks. In 5G NR, the physical resources of the deployed next generation node-B (gNB) can be allocated to several bandwidth parts and configured via numerology to accomplish the data rate and latency requirements of the services.

\begin{figure}[h]
	\centering  
	\begin{center}           
		\includegraphics[width=.50\textwidth]{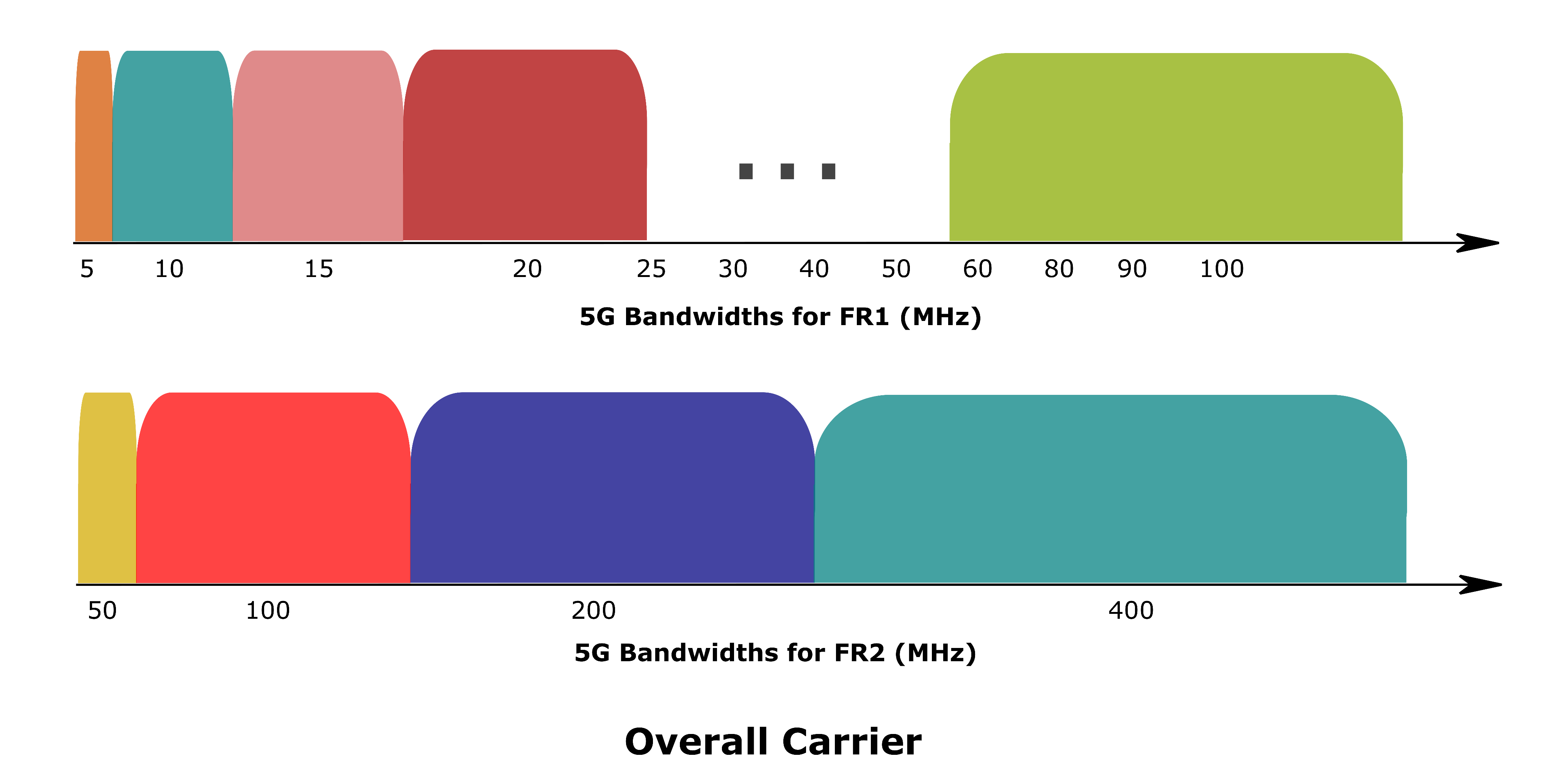}
	\end{center} 
	\caption{Channel bandwidths in 5G for FR1 and FR2}\label{fig:channel_bandwidths}
\end{figure}

Numerology in 5G is defined as the flexibility to use multiple values of sub-carrier spacing $\Delta f$ and is configured through parameter $\mu$ \cite{numerology}. The values of $\Delta f$ depend upon the selected frequency range. 5G NR supports two frequency ranges; the frequency range one is less than 6 GHz (FR1$<$6GHz) and  the frequency range two is greater than 6 GHz (FR2$>$6GHz). The channel bandwidths supported by these two bands are shown in Fig.\ref{fig:channel_bandwidths}.    

To adopt the desired requirements in upcoming 5G spectrum, $3^{rd}$ Generation Partnership Project (3GPPP) has proposed NR with new frequency bands ranging from 1GHz-100 GHz. Moreover, 5G use cases require a different range of parameters such as CP length, sub-carrier spacing ($\Delta f$) and slot duration ($T_{slot}$). These parameters can be configured through numerology to support the desired requirements \cite{numerology} in the access network where the required number of gNBs could be deployed to fulfill capacity and coverage in the target area. Numerology in 5G has been defined as the flexibility to use multiple values of $\Delta f$ and is characterized by parameter $\mu$ as shown in Table \ref{table:trasnmission_num}.

\begin{table}[h]
	\centering
	\caption{5G Transmission Numerologies}\label{table:trasnmission_num}
	\begin{tabular}{lllll}
		\hline\noalign{\smallskip}
		$\mu$ & $\Delta f$  & Cyclic Prefix     &$T_{slot}=1ms/2^\mu$   \\ 		
		\noalign{\smallskip}\hline\noalign{\smallskip}
		0&  15&  Normal &1ms\\ 
		1&  20& Normal  &0.5ms\\ 
		2&  60&  Normal  &0.25ms\\ 
		3&  120&  Normal, Extended   &0.12ms\\ 
		4&  240&  Normal  &0.06ms\\ 
		\noalign{\smallskip}\hline		
		
	\end{tabular}
	
\end{table}

\begin{figure}[h]
	\centering  
	\begin{center}           
		\includegraphics[width=.40\textwidth]{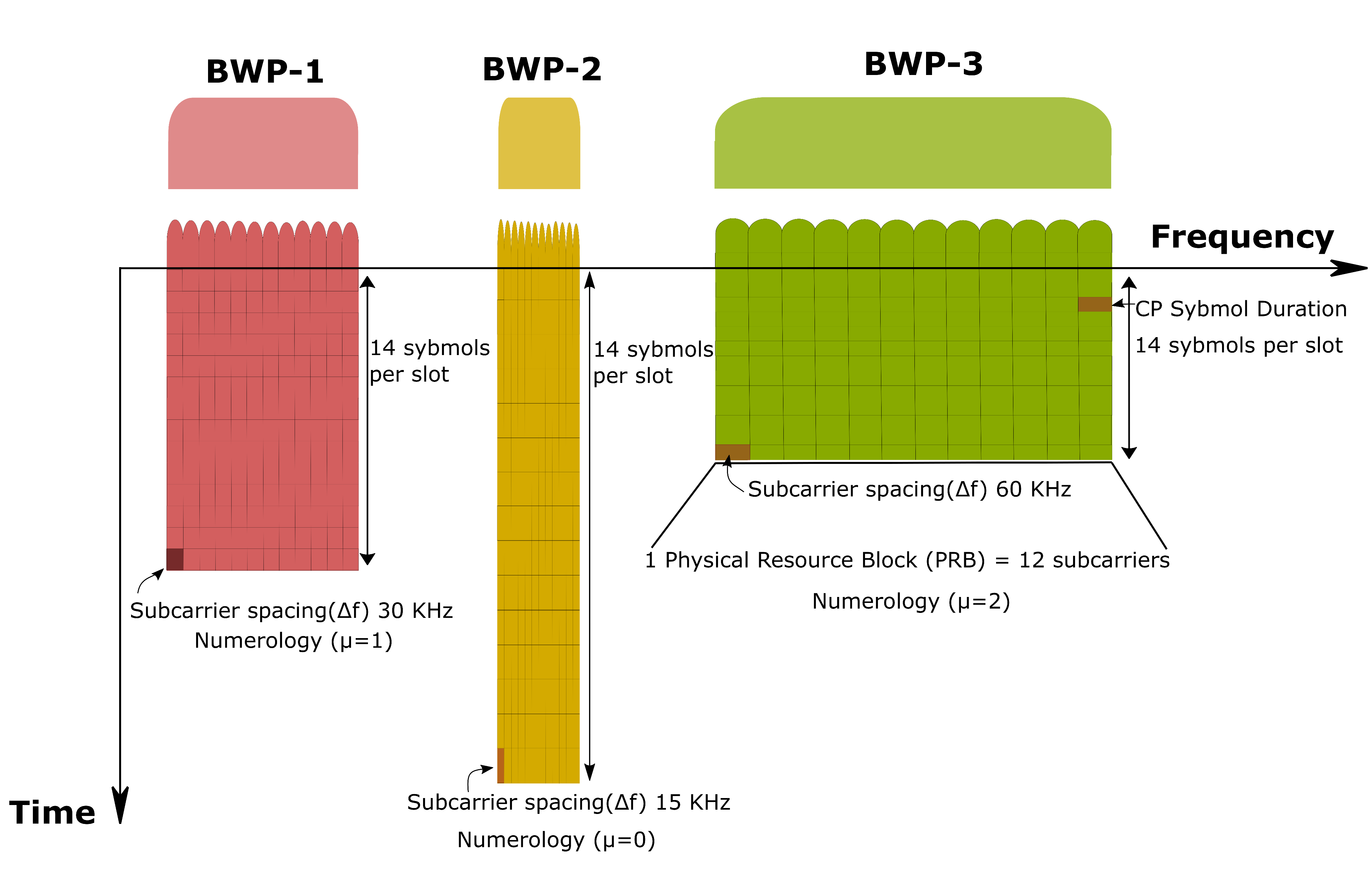}
	\end{center} 
	\caption{Numerology and BWPs for different sub-carrier spacings}\label{fig:NEWRADIO_BWPS}
\end{figure}

A bandwidth part (BWP) refers to an adjacent set of physical resource blocks (PRBs) within overall carrier. These BWPs are configured with different sub-carrier spacings ($\Delta f$) corresponding to a specific value of numerology ($\mu$). In this way, each BWP has different symbol duration and cyclic prefix (CP) length, as shown in Fig.\ref{fig:NEWRADIO_BWPS}.

5G is different from previous generations of cellular networks in the context of versatile use cases or services defined through NR for several peak data rates and latency needs. Therefore offered services are configured by different BWPs through numerology. For instance, considering the BWPs of Fig.\ref{fig:NEWRADIO_BWPS}, services with lower latency requirements can be configured with BWP-1 and BWP-3 as they corresponds to lower orthogonal frequency-division multiplexing (OFMD) symbol durations with $\mu=\left\{1,2\right\}$, respectively. In contrast, BWP-2 with $\mu=0$ has a larger symbol duration which is not suitable for service with lower latency requirements. Compared to LTE, 5G NR for larger values of $\mu \geqslant 0$ can support larger bandwidths with shorter transmission time interval (TTI) which is suitable for mission critical services of lower latency needs. In 5G, the sub-carrier spacing and OFDM symbol duration can be configured with different values of numerology to reduce TTI and access delay.

\section{CONCLUSIONS}
Next-generation 5G network planning demands a paradigm shift from legacy cellular networks, driven by diverse service requirements (eMBB, URLLC, mMTC), cost efficiency, and data-driven precision. This work establishes significant aspects that are considered fundamental to optimized next-generation network planning. First, service-centric design for 5G new radio (NR) features like numerology $(\mu)$ and bandwidth parts (BWPs) are essentials to configure latency, bandwidth, and reliability for heterogeneous use cases. Next, balanced radio network dimensioning (RND) must reconcile coverage and capacity objective, as under-dimensioning casues congestion and over-dimensioning wastes resources. Finally, real-world traffic data extracted via LTE identifiers (TAC, CID), not only enables optimal base station placement in high-density zones but also helps us to plan network with minimal cost per bit.    

\addtolength{\textheight}{-12cm}

\section*{ACKNOWLEDGMENT}

\bibliographystyle{ieeetr}
\bibliography{Bibliography}

\begin{thebibliography}{10}

\bibitem{MonaJou}
M.~{Jaber}, Z.~{Dawy}, N.~{Akl}, and E.~{Yaacoub}, ``Tutorial on {LTE/LTE-A}
  cellular network dimensioning using iterative statistical analysis,'' {\em
  IEEE Communications Surveys Tutorials}, vol.~18, pp.~1355--1383,
  Secondquarter 2016.

\bibitem{Ataufiq}
A.~{Taufique}, M.~{Jaber}, A.~{Imran}, Z.~{Dawy}, and E.~{Yacoub}, ``Planning
  wireless cellular networks of future: {O}utlook, challenges and
  opportunities,'' {\em IEEE Access}, vol.~5, pp.~4821--4845, 2017.

\bibitem{perez2016}
J.~P. {Romero}, O.~{Sallent}, R.~{Ferrús}, and R.~{Agustí}, ``Knowledge-based
  {{5G}} radio access network planning and optimization,'' {\em 2016
  International Symposium on Wireless Communication Systems (ISWCS)},
  pp.~359--365, Sep. 2016.

\bibitem{saxena2017}
N.~{Saxena}, A.~{Roy}, and H.~{Kim}, ``Efficient {{5G}} small cell planning
  with e{MBMS} for optimal demand response in smart grids,'' {\em IEEE
  Transactions on Industrial Informatics}, vol.~13, pp.~1471--1481, June 2017.

\bibitem{tseng2015}
F.~{Tseng}, L.~{Chou}, H.~{Chao}, and J.~{Wang}, ``Ultra-dense small cell
  planning using cognitive radio network toward {{5G}},'' {\em IEEE Wireless
  Communications}, vol.~22, pp.~76--83, December 2015.

\bibitem{udn1}
A.~L. {Rezaabad}, H.~{Beyranvand}, J.~A. {Salehi}, and M.~{Maier},
  ``Ultra-dense {{5G}} small cell deployment for fiber and wireless
  backhaul-aware infrastructures,'' {\em IEEE Transactions on Vehicular
  Technology}, vol.~67, pp.~12231--12243, Dec 2018.

\bibitem{humancentric}
C.~Y. {Lee} and H.~G. {Kang}, ``Cell planning with capacity expansion in mobile
  communications: a tabu search approach,'' {\em IEEE Transactions on Vehicular
  Technology}, vol.~49, pp.~1678--1691, Sep. 2000.

\bibitem{humancentric1}
T.~{Bauschert}, C.~{Büsing}, F.~{D'Andreagiovanni}, A.~M. {C. A. Koster},
  M.~{Kutschka}, and U.~{Steglich}, ``Network planning under demand uncertainty
  with robust optimization,'' {\em IEEE Communications Magazine}, vol.~52,
  pp.~178--185, February 2014.

\bibitem{humancentric2}
H.~{Ghazzai}, E.~{Yaacoub}, M.~{Alouini}, Z.~{Dawy}, and A.~{Abu-Dayya},
  ``Optimized lte cell planning with varying spatial and temporal user
  densities,'' {\em IEEE Transactions on Vehicular Technology}, vol.~65,
  pp.~1575--1589, March 2016.

\bibitem{TS38300}
``{3GPP TS 38.300 V}16.0.0 - technical specification. {NR} and {NG-RAN} overall
  description; stage 2 ({R}elease 16).'' 2019.

\bibitem{qoslte}
T.~{Doumi}, M.~F. {Dolan}, S.~{Tatesh}, A.~{Casati}, G.~{Tsirtsis},
  K.~{Anchan}, and D.~{Flore}, ``Lte for public safety networks,'' {\em IEEE
  Communications Magazine}, vol.~51, pp.~106--112, February 2013.

\bibitem{khasef2016}
M.~{Kashef}, M.~{Ismail}, E.~{Serpedin}, and K.~{Qaraqe}, ``Balanced dynamic
  planning in green heterogeneous cellular networks,'' {\em IEEE Journal on
  Selected Areas in Communications}, vol.~34, pp.~3299--3312, Dec 2016.

\bibitem{maule2018}
M.~{Maule}, D.~{Moltchanov}, P.~{Kustarev}, M.~{Komarov}, S.~{Andreev}, and
  Y.~{Koucheryavy}, ``Delivering fairness and {Q}o{S} guarantees for {LTE/Wi-Fi
  Coexistence Under LAA} operation,'' {\em IEEE Access}, vol.~6,
  pp.~7359--7373, 2018.

\bibitem{qoe1}
K.~{Zheng}, X.~{Zhang}, Q.~{Zheng}, W.~{Xiang}, and L.~{Hanzo},
  ``Quality-of-experience assessment and its application to video services in
  {LTE} networks,'' {\em IEEE Wireless Communications}, vol.~22, pp.~70--78,
  February 2015.

\bibitem{qoe2}
A.~E. {Essaili}, D.~{Schroeder}, D.~{Staehle}, M.~{Shehada}, W.~{Kellerer}, and
  E.~{Steinbach}, ``Quality-of-experience driven adaptive {HTTP} media
  delivery,'' {\em 2013 IEEE International Conference on Communications (ICC)},
  pp.~2480--2485, June 2013.

\bibitem{qoe3}
L.~{Pierucci}, ``The quality of experience perspective toward {{5G}}
  technology,'' {\em IEEE Wireless Communications}, vol.~22, pp.~10--16, August
  2015.

\bibitem{qos2}
Z.~{Niu}, S.~{Zhou}, Y.~{Hua}, Q.~{Zhang}, and D.~{Cao}, ``Energy-aware network
  planning for wireless cellular system with inter-cell cooperation,'' {\em
  IEEE Transactions on Wireless Communications}, vol.~11, pp.~1412--1423, April
  2012.

\bibitem{Yang2016}
Z.~{Yang}, M.~{Chen}, Y.~{Wen}, L.~{Jia}, and Y.~{Zhang}, ``Cell planning based
  on minimized power consumption for {LTE} networks,'' {\em 2016 IEEE Wireless
  Communications and Networking Conference}, pp.~1--6, April 2016.

\bibitem{Gonzalez2016}
D.~{González G.} and J.~{Hämäläinen}, ``Looking at cellular networks
  through canonical domains and conformal mapping,'' {\em IEEE Transactions on
  Wireless Communications}, vol.~15, pp.~3703--3717, May 2016.

\bibitem{Muhnoz2018}
P.~{Muñoz}, O.~{Sallent}, and J.~{Pérez-Romero}, ``Self-dimensioning and
  planning of small cell capacity in multitenant {{5G}} networks,'' {\em IEEE
  Transactions on Vehicular Technology}, vol.~67, pp.~4552--4564, May 2018.

\bibitem{opensignal}
``{O}pen signal mobile analytics.'' 2019. [Online]. Available:
  https://www.opensignal.com/.

\bibitem{nperf}
``Telecommunication coverage maps, {NPERF}.'' 2019. [Online]. Available:
  https://www.nperf.com/.

\bibitem{opencellid}
``Exisiting 4{G} infrastructure from {O}pen{C}elli{D} by {U}niwired {L}abs.''
  2019. [Online]. Available: https://opencellid.org/.

\bibitem{monaBH}
M.~Jaber, M.~A. Imran, R.~Tafazolli, and A.~Tukmanov, ``5g backhaul challenges
  and emerging research directions: A survey,'' {\em IEEE Access}, vol.~4,
  pp.~1743--1766, 2016.

\bibitem{monaBH1}
M.~Jaber, F.~J. Lopez-Martinez, M.~A. Imran, A.~Sutton, A.~Tukmanov, and
  R.~Tafazolli, ``Wireless backhaul: Performance modeling and impact on user
  association for 5g,'' {\em IEEE Transactions on Wireless Communications},
  vol.~17, no.~5, pp.~3095--3110, 2018.

\bibitem{numerology}
A.~A. {Zaidi}, R.~{Baldemair}, H.~{Tullberg}, H.~{Bjorkegren}, L.~{Sundstrom},
  J.~{Medbo}, C.~{Kilinc}, and I.~{Da Silva}, ``Waveform and numerology to
  support {5G} services and requirements,'' {\em IEEE Communications Magazine},
  vol.~54, pp.~90--98, November 2016.

\end{thebibliography}

\end{document}